\documentclass[11pt]{article}
\pdfoutput=1
\usepackage{color}
\usepackage{xcolor}
\usepackage{comment}
\definecolor{darkblue}{rgb}{0.1,0.1,.7}
\usepackage[colorlinks, linkcolor=darkblue, citecolor=darkblue, urlcolor=darkblue, linktocpage,hyperfootnotes=false]{hyperref} 
\usepackage{epsfig}
\usepackage{graphicx}
\usepackage{cite}
\usepackage{amsfonts}
\usepackage{amssymb}
\usepackage{bm}
\usepackage{latexsym}
\usepackage{mathtools}
	\usepackage{amsmath}
	\numberwithin{equation}{section}
	\def\bq{\begin{quote}}
		\def\eq{\end{quote}}
	 at 10truept

	\newcommand{\cala}{{\cal A}}
	\newcommand{\cald}{{\cal D}}
	\newcommand{\cale}{{\cal E}}
	\newcommand{\calf}{{\cal F}}
	\newcommand{\calo}{{\cal O}}

	\newcommand{\cals}{{\cal S}}
	\newcommand{\calt}{{\cal T}}

	\newcommand{\beq}{\begin{equation}}
		\newcommand{\eeq}{\end{equation}}
	\newcommand{\beqa}{\begin{eqnarray}}
		\newcommand{\eeqa}{\end{eqnarray}}
	\newcommand{\bea}{\begin{eqnarray}}
		\newcommand{\eea}{\end{eqnarray}}
	
	\newcommand{\hf}{\frac{1}{2}}

	\def\lesssim{~\mbox{\raisebox{-.6ex}{$\stackrel{<}{\sim}$}}~}
	\def\roughly#1{\raise.3ex\hbox{$#1$\kern-.75em\lower1ex\hbox{$\sim$}}}

\newcommand\olm{{\omega l m}}

\newcommand\surf{\cala}
\newcommand\tube{\cals}
\newcommand\rup{{\rm up}}
\newcommand\rdown{{\rm down}}
\newcommand\rin{{\rm in}}
\newcommand\rout{{\rm out}}
\newcommand\rsc{{\rm sc}}
\graphicspath{{images/}}

\begin{document}

		\thispagestyle{empty}
		\begin{titlepage}
			\bigskip
			
			\bigskip\bigskip

			\bigskip
			
			\begin{center}
				{\Large \bf {Deriving effective descriptions  and signal predictions for dynamical gravitational systems}}
				\bigskip
				\bigskip
			\end{center}

			\begin{center}

				\rm { Steven B. Giddings\footnote{\texttt{giddings@ucsb.edu}} and Madhur Mehta\footnote{\texttt{madhurtime17@ucsb.edu}}}
				\bigskip \rm
				\bigskip
				
				{Department of Physics, University of California, Santa Barbara, CA 93106, USA}  \\
				\rm
				
				\bigskip \rm
				\bigskip
				
				\rm

				\bigskip
				\bigskip

			\end{center}

			\vspace{3cm}
			\begin{abstract}
			We investigate top-down derivations of effective descriptions of radiation from gravitational systems such as binaries. With a specified cutoff prescription, one can derive worldline effective field theories, although the associated cutoff dependence can complicate their formulation. We investigate a related effective approach in which the dynamics are parameterized by an action on the boundaries of cavities enclosing the individual bodies. We present examples of such cavity descriptions for black holes and for simple models of modified black hole behavior. We also show how cavity effective descriptions connect to observable quantities including Love numbers and aspects of wave profiles, such as the accumulated phase shifts of emitted signals. A primary motivation is to develop a systematic framework for inferring how modifications of classical black hole behavior affect gravitational wave signals. Such modifications may be motivated by the requirement that black hole evolution be consistent with quantum mechanics, or by other models of new black hole behavior. Accumulated phase shifts during the inspiral have, in particular, been argued to provide sensitivity to small new effects. To illustrate the basic principles clearly, we develop the formalism primarily for scalar radiation, and then outline its extension to gravitational perturbations.
			
		
				\medskip
				\noindent
			\end{abstract}
			\bigskip \bigskip \bigskip 
			
		\end{titlepage}
	
	\tableofcontents
	
	\newpage
	
\section{Introduction}
	
The new era of gravitational wave (GW) astronomy has opened a window to increasingly precise tests of gravitational physics~\cite{Abbott:2016blz,Abbott:2016nmj,LIGOScientific:2019fpa,LIGOScientific:2020tif,LIGOScientific:2021sio}.  A key aspect of this is to improve prediction of waveforms, and contributions to them from underlying physics, that can be compared with GW observations; refinement of these calculations, improved observational methods, and statistical power from the growing number of observations are all contributing to this increasing precision. A central question is whether the GW observations associated to black hole (BH) mergers are in fact accurately described via a general relativistic treatment of classical black holes, or whether there are new effects contributing to the signals.  BH physics poses a test of general relativity in the extreme limit of strong gravitation.
There are numerous proposals for new physics that could be evident here, ranging from classical modifications of the action or field content, to quantum modifications necessary~\cite{Hawking:1975vcx,Hawking:1976ra,Page:1993wv} for a consistent quantum description of BHs (see, {\it e.g.}, \cite{Mathur:2005zp,Giddings:2017mym,PSSY,AHMST,MaMa,Almheiri:2020cfm,UCNVU}).

Extracting signal predictions from general relativity (GR) or some modification to it is of course quite challenging, in large part due to the nonlinearity of GR.  A traditional approach has used numerical methods  including numerical relativity~\cite{Pretorius:2005gq,Campanelli:2005dd,Baker:2005vv,Lehner:2014asa}.  However, at least in certain regimes, such as the initial inspiral of a binary BH system, another approach has been developed with  increasingly refined predictions.  The basic idea is that in regimes where velocities are low and wavelengths of emitted radiation are correspondingly long, one can treat the physics in expansions about a limit where BHs are point particles, and the dynamics is that of Newtonian gravity~\cite{Blanchet:2013haa,Buonanno:1998gg,Buonanno:2000ef,Damour:2011fu}.  The first of these is associated with what has been called worldline effective field theory (EFT)~\cite{Goldberger:2004jt,Goldberger:2005cd,Goldberger:2009qd,Porto:2005ac,Porto:2016pyg,Foffa:2013qca,Levi:2018nxp}, and the second is the post-Newtonian expansion~\cite{Blanchet:2013haa,Foffa:2013qca,Levi:2018nxp}.  These combined approaches have led to increasingly precise predictions of GW templates and comparison with observation.

In addition to refining these methods, one would like to explore their extension to treating possible modifications to classical general relativistic BH behavior, due to either new classical or quantum effects.  One challenge for this appears to arise from an incomplete understanding of the worldline EFT approach.  Specifically, as traditionally developed in~\cite{Goldberger:2004jt,Goldberger:2005cd,Goldberger:2009qd}, it postulates long-distance actions, guided by symmetry and other considerations, and then matches their predictions to those of a microscopic theory, such as the more complete GR description.  Such a ``bottom-up" approach has been used for other EFTs, such as the Fermi theory of weak interactions, the Standard Model, and its extensions.  However, in many such other examples of EFT, we can alternately begin with the more fundamental theory and {\it derive} the EFT in an approximation, such as in the case where the four-fermion interaction is derived as a low energy approximation to the full electroweak description involving $W$ or $Z$ exchange.  Developing such a ``top-down" approach to an effective theory describing GW emission 
seems particularly relevant for investigating possible modifications of GR.  Specifically, if BHs are described by some more basic dynamics that is a modification of GR ({\it e.g.} as in \cite{Mathur:2005zp,Giddings:2017mym,PSSY,AHMST,MaMa,Almheiri:2020cfm,UCNVU}), we would like to have a way to derive the longer-distance consequences of that, for example on GW signals, through an effective theory approach.

In this paper we begin investigation of this question of top-down derivation of long-distance effective descriptions, and their use in both parameterizing modifications to classical BH behavior, and the effect of those modifications on GW signals.  To understand the basic approach and principles, a good starting point is the simple model of scalar field perturbations on a non-spinning BH background.
We will investigate the connection of such a microscopic theory to long-wavelength effective descriptions; treatment of gravitational perturbations and spinning BHs are then, as we will discuss, extensions of this basic approach with some additional technical machinery.  We will specifically also begin to estimate effects of simple models for deviations from classical BH behavior, and their parameterization in an effective description, and resulting effects on GW signals.

Our starting point, in section two, is a generating functional parameterization of scalar field dynamics in a BH background.  The corresponding functional integral can be factorized across a timelike tube containing the BH.  The dynamics of the BH and its excitations interior to the tube can then be parameterized using an effective action on the tube boundary, in what one might describe as a ``cavity description" of the physics.  One might expect such a description to be a starting point for the derivation of worldline EFT.  However, there are some puzzles in directly connecting worldline EFT with more fundamental dynamics, which we outline.  These suggest that the connection may only be precise when one has a definite cutoff prescription for the worldline EFT.  We investigate possible versions of such a cutoff, in the context of the cavity picture.  One can derive a worldline EFT, and a long-wavelength expansion and approximation, in this fashion, but there is nontrivial cutoff dependence that must cancel between the regulated point particle description we give, and the terms in the effective action parameterizing deviations from the point particle description, {\it e.g.} incorporating BH behavior such as absorption.

The fact that this cutoff dependence should cancel in the cavity description suggests that this description is a useful one for parameterizing the physics, and we further develop this and its long-wavelength limit  in the rest of the paper.

Specifically, section three develops cavity effective descriptions for examples of static, spherically-symmetric geometries: classical Schwarzschild, and modifications involving models of new interactions either microscopically close to the horizon, or in the BH atmosphere at distances of order the Schwarzschild radius from the horizon~\cite{Cardoso:2016rao,Abedi:2016hgu,Cardoso:2017cqb,Cardoso:2019rvt,Mark:2017dnq,Maggio:2017ivp,Maggio:2019zyv}\cite{Giddings:2017mym,UCNVU}.  These illustrate basic features of the approach.

Section four then turns to applications of cavity effective descriptions.  In a first application, one can relate a cavity description to a parameterization of the physics in terms of the amplitude of the outgoing field signal, in a scattering description developed in \cite{FrGi}.  The cavity description determines the field amplitudes, and vice versa.  In a second application, we focus on a particular aspect of GW signals: GWs carry away orbital binding energy, and so any modification to such signals can alter the rate at which the orbit decays.  The many orbits of inspiral then acts as a possible amplifier that turns small corrections into more significant corrections to the overall GW phase, as has been investigated in \cite{Flanagan:2007ix,Hinderer:2007mb,Damour:2009wj,Datta:2019epe,Maggio:2021uge,Sago:2022bbj}\cite{FrGi,Seymour:2024orl,CZI}.  We outline an approach to calculating such corrections and their effect on the phase from the cavity approach, in the scalar model.

Section five outlines various directions for future work.  One is fuller treatment of spin two metric perturbations.  Another is extending these effective descriptions to spinning BHs.  We also anticipate further developing models for modifications of BH interactions, due to the need for consistency with quantum mechanics, as with nonviolent unitarization (see {\it e.g.} \cite{Giddings:2017mym,UCNVU}, and references therein), and using the effective methods of this paper to parameterize their effects on GW signals.  Another anticipated development is more direct incorporation of BH position and velocities into such cavity effective descriptions.  Also, there is nontrivial running of these effective descriptions with scale, whose renormalization group behavior warrants investigation.

Two appendices give a more complete description of the connection to worldline EFT, and of calculation of and scale behavior for Love numbers in a cavity approach.
	
\section{Derivation of effective descriptions}

\subsection{Generating functional and cavity description}

Our first goal is to describe the mutual effects of a classical BH, or of a more general compact quantum object, on its surrounding fields.  Combining such a description for two such objects coupled to the gravitational field then provides a way to predict the gravitational wave signal of their inspiral, and then to include any new effects departing from classical BH predictions.  For a simple illustration of our approach, we first describe couplings to a scalar field, and then describe how these are modified in the case of higher spin fields such as the gravitational field.

Specifically, we consider a localized object with metric $g_{\mu\nu}$; in this paper we will primarily focus on the case with zero angular momentum, but will be general for now.  We can describe its interactions with a scalar field $\phi$ in terms of the generating functional with general source $J$, which we can represent  as a functional integral,
\beq\label{genfunct}
Z[J] = \int \cald \phi\ e^{iS[\phi] + i \int dV_4 J(x)\phi(x)}\ .
\eeq
Here $S[\phi]$ is the scalar action, for example the free action
\beq\label{phiact}
S[\phi]=-\hf \int dV_4 \left[ (\nabla\phi)^2 + m^2\phi^2\right]\ ,
\eeq
with four-volume element $dV_4= \sqrt{|g|} d^4x$.  To describe the coupling of the object to $\phi$, we choose a surface $\surf$ surrounding the object, whose evolution defines a worldtube $\tube$ surrounding the worldline of the object; we take the support of the source to be outside this worldtube.  Then, the generating functional can be trivially decomposed into internal and external parts, denoted with $<$ and $>$, 
\beq\label{genfeq}
Z[J] =   \int _\tube D\varphi  \int_<^{\phi_{\vert\tube}=\varphi} \cald \phi\  e^{iS[\phi] }\int_>^{\phi_{\vert\tube}=\varphi}  \cald \phi\ e^{iS[\phi] + i \int dV_4 J(x)\phi(x)} \
\eeq
together with a functional integral over a common boundary value $\varphi$ on $\tube$.  
We can rewrite this as
\beq\label{zwbdy}
Z[J] = \int_>\cald\phi\ e^{iS[\phi] + i \int dV_4 J(x)\phi(x) + i S_\partial[\varphi]}
\eeq
where the integral now includes one over free boundary values $\phi_{\vert\tube}=\varphi$, and
where we think of the functional integral over fields inside $\tube$ as defining a ``boundary action,"
\beq\label{bfix}
e^{ i S_\partial[\varphi]}=\int_<^{\phi_{\vert\tube}=\varphi}\cald\phi\ e^{iS[\phi]}\ .
\eeq

\begin{figure}[t!]
 	\begin{center}
 		\includegraphics[width=1.0\textwidth]{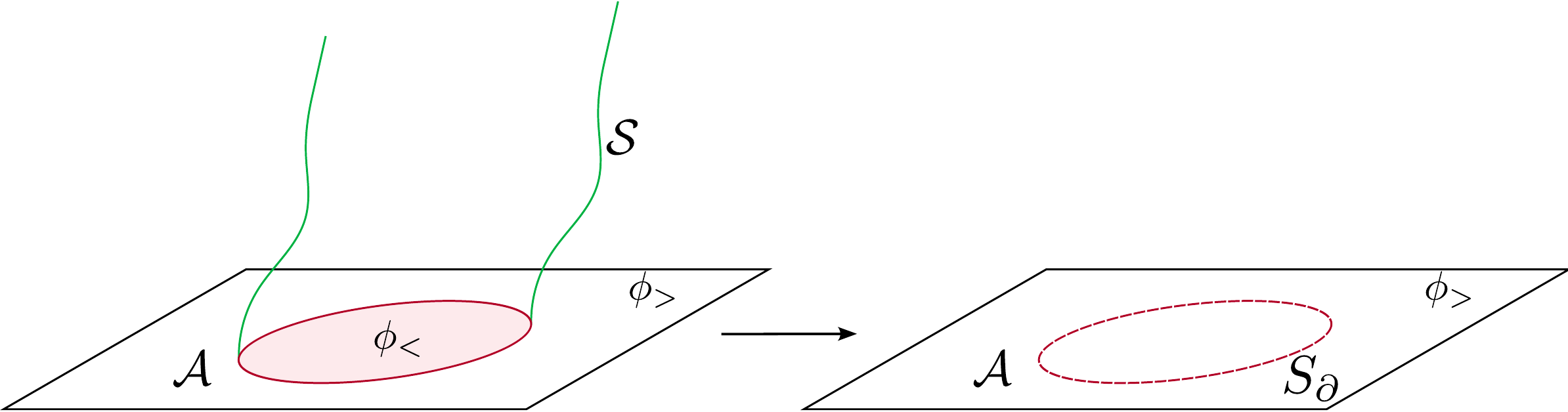} 
 		\caption{Schematic of the cavity description.  A region inside an area $\surf$, sweeping out a worldtube $\tube$, has its dynamics captured by a boundary action $S_\partial$ which can be found for example by factorization of the functional integral giving the generating functional.}
 		\label{fig:cavity}
 	\end{center}
 \end{figure}

The  action $S_\partial$ summarizes the interactions of the object inside $\tube$ with the fields outside.  The resulting functional integral is over the exterior of $\tube$, with the interior removed; for this reason we refer to this as a ``cavity" description of the dynamics.  When considering a  $J(x)$ that sources excitations that are long wavelength as compared to the size $\rho$ of the cavity, we anticipate that the action $S_\partial$ should be well approximated by one involving local operators, and that this is how to connect to the worldline effective field theory descriptions of \cite{Goldberger:2004jt,Goldberger:2005cd,Goldberger:2009qd,Porto:2005ac,Porto:2016pyg,Foffa:2013qca,Levi:2018nxp}. We will return to this question below.

We first investigate $S_\partial$ more closely.  It can be written in terms of the local field $\varphi$ on $\tube$ by introducing a functional delta function,
\beq
\Delta[\phi_{|\tube}-\varphi] = \int D\alpha\ e^{i \oint d\tube \alpha(x)[\phi(x)-\varphi(x)]}\ ,
\eeq
where $d\tube$ is the area element for integration over $\tube$, resulting in
\beq\label{afix}
e^{iS_\partial[\varphi]} = \int D\alpha\ e^{-i\oint d\tube \alpha(x) \varphi(x)} \int_< \cald\phi\ e^{iS[\phi] +i \oint d\tube \alpha(x)\phi(x)} \ 
\eeq
with free boundary conditions for $\phi$ at $\tube$.
One can think of the first exponential as describing a local coupling of the exterior field operator $\varphi(x)$ to the internal degrees of freedom, as in the worldline EFT descriptions of \cite{Goldberger:2004jt,Goldberger:2005cd}, although here the field couples to a conjugate $\alpha$ to the internal field $\phi$.  

To continue, one must evaluate the functional integrals \eqref{bfix} or \eqref{afix}.  For a general interacting theory, the corresponding functional integrals over $\phi$ produce a correspondingly complicated functional of $\varphi$ or $\alpha$.  This simplifies to quadratic dependence in the case of a quadratic action for the field, such as the minimal action \eqref{phiact}.  In this latter case, beginning with \eqref{bfix}, we can write the integration variable as
\beq
\phi = \bar\phi + \tilde\phi
\eeq
where $\bar\phi$ satisfies the classical equation of motion and the boundary condition $\bar\phi_{|\tube}=\varphi$, and where the fluctuation $\tilde\phi$ satisfies $\tilde \phi_{|\tube}=0$.  Then, after integration by parts, we find
\beq
e^{iS_\partial[\varphi]} = \exp\left\{-\frac{i}{2}\oint d\tube \varphi \nabla_n \bar\phi + iS_\partial[0]\right\}
\eeq
where $\nabla_n$ is the normal derivative to $\tube$  in the {\it outward} direction.  Then the equation of motion for $\bar \phi$ may be used to rewrite this derivative in terms of the boundary value of the field and a ``response function" $K(x,x')$ given by a Green function,
\beq\label{respdef}
\nabla_n\bar\phi(x') = \oint d\tube \varphi(x) K(x,x') \ ,
\eeq
resulting in
\beq\label{sbdy}
S_\partial[\varphi]= -\hf \oint\oint d\tube d\tube' \varphi(x) K(x,x') \varphi(x') + S_\partial[0]\ .
\eeq
Explicitly, $K$ can be written
\beq\label{kfree}
K(x,x') = \nabla_n\nabla_{n'}G_D(x,x') = \langle \nabla_n\phi(x) \nabla_{n'}\phi(x')\rangle_D\ ,
\eeq
where $G_D$ is the Green function with Dirichlet boundary conditions for $\phi$.  Alternately, by working with the conjugate variable $\alpha$ in \eqref{afix}, one may write $K$ in terms of an {\it inverse} to the Neumann Green function for $\phi$.\footnote{To elaborate, note that in the case of quadratic action, the functional integral over $\phi$ in \eqref{afix} may be thought of as computing a two-point function.  Then, the integral over $\alpha$ gives an expression involving its inverse.}

In summary, the internal dynamics of the object is captured by including $S_\partial$ in a cavity description \eqref{zwbdy}; in the case of a quadratic internal action $S_\partial$,  \eqref{sbdy},  is determined by the two-point function of the internal field, with the explicit result \eqref{kfree} for a free scalar field.

This structure straightforwardly generalizes to other fields, for example with spin.  Consider a collection $\phi_A$ of fields, which could for example be the components of a field with spin, and corresponding sources $J_A$, giving a generating functional generalizing \eqref{genfunct}.  These could be, for example, components of the metric perturbation $h_{\mu\nu}$.  The generating functional may be factorized, following the preceding steps, resulting in a cavity description with a boundary action $S_\partial[\varphi_A]$, and also rewritten in terms of conjugate variables $\alpha_A$ as in \eqref{afix}.  For a quadratic action, integration by parts results in a linear field equation with a differential operator $L$, and one expects the boundary action to be determined in terms of Green's functions for that operator.

We also expect to be able to use \eqref{sbdy} to describe more general dynamics than governed by local propagation in a classical metric, if it yields a linear response, and more general nonlinear dynamics to be described by more general functional dependence in $S_\partial$.  This gives a natural way to incorporate a wide class of scattering behaviors of more general objects than classical black holes, which we begin to explore below.

%

\subsection{Towards worldline effective field theory: point particle descriptions}\label{PPsec}

We next compare the preceding exact description of the dynamics with that of the worldline effective field theory approach~\cite{Goldberger:2004jt,Goldberger:2005cd,Goldberger:2009qd,Porto:2005ac,Porto:2016pyg,Foffa:2013qca,Levi:2018nxp}.  

The worldline EFT approach has been widely used in the case of metric perturbations, to describe gravitational wave signals.  In contexts where metric perturbations are expected to be long wavelength, such as binary black hole inspiral, the expectation has been that the leading description of the dynamics of black holes  is given by a point particle action coupled to the Einstein-Hilbert action,
\beq\label{ppact}
S=S_{pp}+S_{EH} = -M\int d\tau \sqrt{-g_{\mu\nu}(X(\tau)) \frac{dX^\mu}{d\tau}\frac{dX^\nu}{d\tau}} + \frac{1}{16\pi G} \int d^4x \sqrt{|g|} R\ .
\eeq
Subleading terms describing internal structure of the BH are then postulated to be described by higher dimension couplings~\cite{Goldberger:2004jt,Goldberger:2005cd,Goldberger:2009qd}, such as
\beq\label{sublead}
S_2 = -\int d\tau\left( Q^E_{ab} E^{ab} +Q^B_{ab} B^{ab} \right)
\eeq
where $E^{ab}$ and $B^{ab}$ are electric/magnetic components of the Weyl tensor (see \cite{Goldberger:2005cd}), and $Q^E_{ab}$, $Q^B_{ab}$are effective multipole moments.
Additional higher dimension terms are expected to describe further subleading dynamics.  Coefficients in these terms are then fixed by  matching calculations to ones carried out in the exact BH description.

We might expect to be able to {\it derive} such an effective description, at long wavelengths, from the kind of exact description given in the preceding section.  We have in mind a derivation analogous to the derivation of the four-fermion weak interaction as a leading order term when one expands the intermediate W-boson propagator for small momentum, or more general discussions of effective field theory such as in \cite{Polchinski:1983gv}.

However, we encounter two related puzzles.

Starting with \eqref{zwbdy} and \eqref{afix}, we find expressions with a similar structure to the EFT.  Define
\beq
e^{iW[\alpha]} =  \int_< \cald\phi\ e^{iS[\phi] +i \oint d\tube \alpha(x)\phi(x)}\ ,
\eeq
so that \eqref{afix} and \eqref{zwbdy} together become
\beq\label{zrew}
Z[J]=\int D\alpha\ e^{iW[\alpha]} \int_>\cald\phi\ e^{iS[\phi] + i \int dV_4 J(x)\phi(x) -i\oint d\tube \alpha(x) \varphi(x)}\ .
\eeq
In the case where this is generalized for metric perturbations, we might expect to be able to describe the point particle action in the first term of \eqref{ppact} through the ``internal" action $W[\alpha]$.  The couplings in the last term of the exponential in \eqref{zrew} appear analogous to those of \eqref{sublead}, although the couplings are to the dual internal variable $\alpha$.  But a possibly more puzzling aspect is that the coupling is to the fields on the worldtube $\tube$, not to local operators at the particle's ``location."  To connect the two, one would like to ``fill in" the interior of the tube.

This leads to the second puzzle. Suppose \eqref{ppact} is taken to be the leading action.  Solutions to this action are {\it not} point particles -- they are black holes, of mass $M$.  So, the leading action already contains  the physics that the corrections \eqref{sublead} were intended to capture.

If we wanted \eqref{sublead} and further terms to summarize the description of non-trivial BH dynamics, such as absorption, it seems we should start with an action that gives solutions that behave more like actual point particles, and not black holes to begin with.  

These puzzles appear to call for us to supplement the description of \eqref{ppact} and \eqref{sublead} with some cutoff prescription.

To explore ways to do this, we return to the simpler but illustrative case of scalar field perturbations, which can then again be generalized. Specifically, define the generating functional \eqref{genfunct} for the true configuration, {\it e.g.} given by the metric $g_{\mu\nu}$; for concreteness, this could be imagined to be the Schwarzschild metric for a BH with radius $R=2M$ (working here and for much of the following with units $G=1$)
\beq\label{schw}
ds^2 = -f(r)dt^2 + \frac{1}{f(r)} dr^2 + r^2 d\Omega^2\ ,
\eeq
with
\beq
f(r) = 1-\frac{R}{r}\ .
\eeq
We imagine that we modify the dynamics \eqref{ppact} by including an explicit cutoff, or for example some matter distribution, to produce a different background metric $\bar g_{\mu\nu}$ with no horizon, and with a regular origin rather than a singularity at $r=0$; {\it e.g.} consider a solution of the form
\beq\label{regmet}
d\bar s^2 = -f_{\bar \rho }(r) dt^2 + g_{\bar \rho}(r) dr^2 + r^2 d\Omega^2\ ,
\eeq
which agrees with \eqref{schw} for $r>\bar \rho>R$, but which removes the horizon and smooths the origin inside $\bar\rho$, as in the case of the 
metric of a star.\footnote{We will consider an alternate description based on reflecting boundary conditions below.  Another example in an analogous context is the shell metric of \cite{Kosmopoulos:2025rfj}.  After posting the first version of this paper, we became aware of \cite{Cohen:2026ovm}, which likewise uses a shell metric, in the limit of large radius shell, to match to worldline EFT at zero frequency.}  Then the generating functional $\bar Z$ defined as in \eqref{genfunct} but with the metric $\bar g_{\mu\nu}$ does not describe BH properties -- but it does depend on the details of the cutoff prescription, and in particular $\bar \rho$.  A cavity description for its dynamics may be analogously introduced by defining the cavity boundary outside the region where  $ g_{\mu\nu}$
 and  $\bar g_{\mu\nu}$ differ, for example in the BH case at $r=\rho>\bar \rho$.  In particular, this defines a boundary action $\bar S_\partial[\varphi]$ with a response function $\bar K$.  
 
 \begin{figure}[t!]
 	\begin{center}
 		\includegraphics[width=1.0\textwidth]{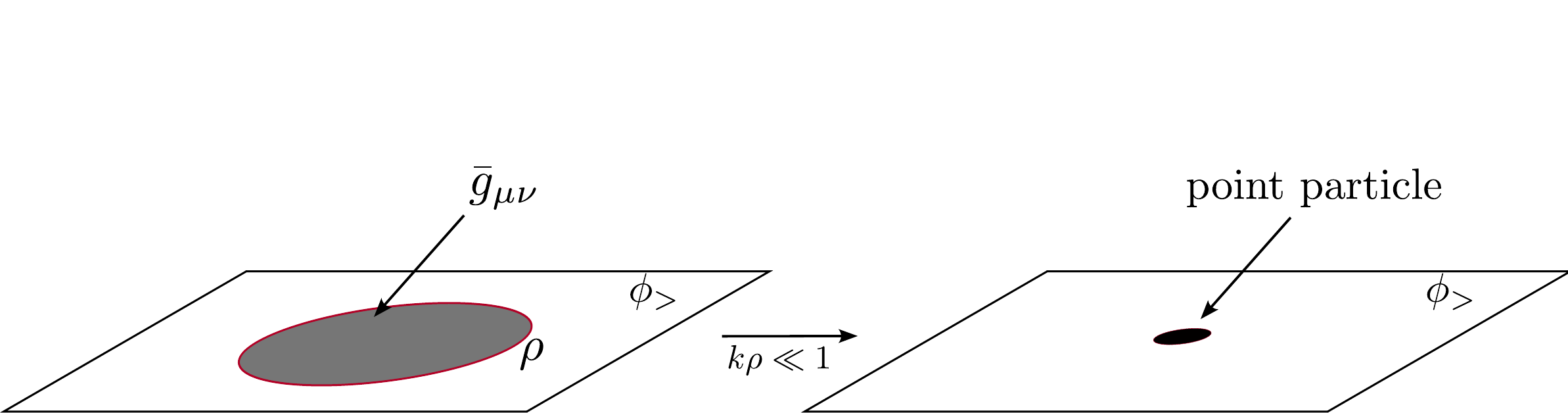} 
 		\caption{For purposes of describing coupling to long-wavelength modes, we expect to be able to replace dynamics inside a surface of radius $\rho$ by the action of an effective ``point particle," with suitable couplings.  For an internal point particle description without black hole features, one can consider a regulated solution with a metric $\bar g_{\mu\nu}$ that is regular at the origin.}
 		\label{fig:PP}
 	\end{center}
 \end{figure}

Black hole or other dynamics of a fundamental description may then be treated as a deviation from this regulated ``point particle" dynamics.  To do so, trivially rewrite \eqref{zwbdy} as
\beq
Z[J] = \int_>\cald\phi\ e^{iS[\phi] + i \int dV_4 J(x)\phi(x) + i \Delta S_\partial[\varphi] + i \bar S_\partial[\varphi] }\ ,
\eeq
with the definition
\beq
\Delta S_\partial[\varphi] = S_\partial[\varphi] -\bar S_\partial[\varphi] \ .
\eeq
Then use the identification 
\beq
 \phi(x)  \leftrightarrow \frac{1}{i} \frac{\delta}{\delta J(x)} 
 \eeq
 inside the functional integral to rewrite this as
 \beq
 Z[J] = \exp\left\{i\Delta S_\partial\left[\frac{1}{i} \frac{\delta}{\delta J(x)}\right]\right\} \int_>\cald\phi\ e^{iS[\phi] + i \int dV_4 J(x)\phi(x) + i \bar S_\partial[\varphi] }
=\exp\left\{i\Delta S_\partial\left[\frac{1}{i} \frac{\delta}{\delta J(x)}\right]\right\} \bar Z[J]\ .
 \eeq
The backgound $\bar Z[J]$ can then be written in the original form \eqref{genfunct}, and the derivatives moved back into the integral, to give a functional integral encompassing the origin,
\beq\label{genfp}
 Z[J] = \int \cald\phi e^{i\bar S[\phi] + i \int d\bar V_4 J(x)\phi(x) + i\Delta S_\partial[\varphi]}\ ,
 \eeq
 where barred quantities are calculated with metric $\bar g_{\mu\nu}$.  
 
 The expression $\Delta S_\partial$ still involves an integral over the worldtube defining the original matching boundary, which in the simple example is at $\rho$.  However, we expect that we can expand $\Delta S_\partial$ in terms of local operators, now that we are working in a description where $r=0$ is a regular point, and that this is a useful expansion when the sources have wavelengths long as compared to the size of the tube.  
 We outline this construction in Appendix~\ref{KtoEFT}.
  This gives an approach to deriving an effective field theory description of the complete underlying dynamics, in terms of point particle dynamics together with corrections involving higher-dimension operators. 
 
 While this makes contact with the worldline EFT approach, it apparently introduces added complexity.  The background $\bar Z[J]$ and correction $\Delta S_\partial$ both depend on the cutoff metric $\bar g_{\mu\nu}$ and in particular on the scale $\bar \rho$.  In this description, there are different EFTs, associated with the different regulators.  While we could study expressions with such explicit dependence, that dependence should cancel in the final result $Z[J]$.  
 
 This suggests that while worldline EFTs can be derived in a precise way from more complete underlying dynamics,
these introduce additional complications arising from the need for a necessarily ad hoc regulator of the short-distance behavior.  We could pursue this description further, but instead turn to developing an approach based on this lesson.  The simple expression without these canceling dependencies on the regulator is the original cavity description of \eqref{zwbdy}.  

This also suggests a type of renormalization group, in which we vary the scale $\bar \rho$ associated with the smooth regulator.  Both $\bar Z[J]$ and $\Delta S_\partial$ then have non-trivial dependence on $\bar \rho$, but that dependence cancels in $Z[J]$ which behaves like an RG invariant.  We will discuss a different, but similar, RG connected with the cavity description, below.

\subsection{Long wavelength dynamics and cavity effective description}\label{CED}

We next turn to exploring the cavity effective description of \eqref{zwbdy} and how it can give a simple description of aspects of the dynamics.  Coupling of the field  $\phi$ to the internal dynamics is summarized by $S_\partial$.  Considering boundary variations in \eqref{zwbdy} shows that $S_\partial$ imposes a boundary condition 
\beq\label{cbc}
\nabla_n \phi_{\vert \tube} = -\frac{\delta S_\partial}{\delta \varphi} \ ,
\eeq
on the quantum field, where again $n$ denotes the normal in the outward direction.  
In the simple case of a non-interacting internal theory \eqref{sbdy}, this takes the form
\beq\label{bdfixni}
\nabla_n \phi(x)_{\vert \tube} =\oint d\tube' K(x,x')\phi(x')\ .
\eeq
We expect these expressions to usefully describe the coupling of fields outside the object to its internal dynamics, and to summarize for example energetics, absorption, and reflection from the object.

While such a description could also be formulated for a spinning object, we will largely focus on the case with a static and spherically symmetric metric, to simplify development of the basic ideas.  So, outside $\tube$, which we take to be at $r=\rho$, the metric has the Schwarzschild form 
\eqref{schw}.  The symmetries then simplify $K$, which can be expanded
\beq\label{KSexp}
K(x,x') = \int \frac{d\omega}{2\pi} \sum_{lm} e^{-i\omega(t-t')} Y_{lm}(\Omega) Y_{lm}^*(\Omega') K_l(\omega,\rho)\ .
\eeq

We expect further simplification in the low-frequency limit, analogous to the EFT simplification arising from the derivative expansion.  Specifically, for $l\lesssim \omega \rho$, we expect the propagation of external excitations to the cavity wall to be suppressed by the angular momentum barrier, and so in particular for $\omega \rho\ll 1$, the lowest angular momenta contributions to dominate.

These, and other features of the cavity description, can be usefully illustrated by considering some simple examples.

\section{Cavity effective description -- examples}\label{csec}

Cavity effective descriptions are readily illustrated in the case of internal dynamics corresponding to propagation in a Schwarzschild metric, or in some proposed modifications to that propagation.  We consider examples where the metric outside $r=\rho$ is Schwarzschild \eqref{schw}.  We first summarize some of the basics of propagation in Schwarzschild, and then consider specific examples of cavity descriptions.

The preceding effective description can be readily illustrated in the case of an interior theory consisting of propagation in a Schwarzschild metric, or in simple modifications to that propagation.  In either case we take the exterior of the matching surface at $r=\rho$ to be the Schwarzschild form \eqref{schw}.  Solutions of the Klein-Gordon equation
\beq\label{kgeq}
(\square-m^2)\phi =0
\eeq
in this background can be expanded in a basis of solutions of the form
\beq\label{schbas}
\phi_{\omega l m}(t,r,\Omega) = e^{-i\omega t} \phi_{\omega l}(r) Y_{lm}(\Omega)=e^{-i\omega t} \frac{u_{\omega l}(r)}{r} Y_{lm}(\Omega)\ ,
\eeq
with radial wavefunctions $\phi_{\omega l}(r)= u_{\omega l}(r)/r$.
Introducing the tortoise coordinate
\beq
\frac{d r_*}{d r} = \left(1-\frac{R}{r}\right)^{-1} 
\eeq
simplifies the radial equation to the form
\beq\label{radeqnt}
\left[\partial_{r_*}^2 +\omega^2 - V_l^{\rm RW}\left(r\right) \right] u_{\omega l}\left(r\right)=0\ ,
\eeq
with Regge-Wheeler potential
\beq\label{rwpot}
V_l^{\rm RW}(r) = \left(1-\frac{R}{r}\right)\left[\frac{l(l+1)}{r^2}+\frac{R}{r^3}+m^2\right]\ .
\eeq
Since our ultimate focus will be on gravitational radiation, the remainder of the paper will focus on the analogous  massless example, $m=0$.

For a complete basis, one needs waves propagating in both directions.  One such basis is specified at $r,r_*\rightarrow \infty$ as
\beq\label{updown}
u^\rup_{\omega l} \rightarrow e^{i\omega r_*}\ ,\ u^\rdown_{\omega l} \rightarrow e^{-i\omega r_*}\ .
\eeq
An alternate basis is specified by modes that are ingoing or outgoing at the horizon, $r\rightarrow R$, $r_*\rightarrow-\infty$,
\beq\label{inout}
u^\rin_{\omega l}  \rightarrow A e^{-i\omega r_*}\ ,\ u^\rout_{\omega l}  \rightarrow B e^{i\omega r_*}\ .
\eeq
The two bases are related through transmission/reflection coefficients in the Regge-Wheeler potential, for example as
\beq
u^\rin_{\omega l}  = u^\rdown_{\omega l}  + R_{\omega l} u^\rup_{\omega l} \ ,
\eeq
with reflection coefficient $R_{\omega l} $; this fixes the normalizations $A$ and $B$.  The solutions \eqref{updown} and \eqref{inout} can be written explicitly in
terms of confluent Heun functions, or equivalently represented by standard
series methods such as the MST expansion~\cite{Ronveaux:1995,Leaver:1986vnb,Fiziev:2005ki,Mano:1996vt,Sasaki:2003xr}.  Note also that $u^\rdown_{\omega l}(r)= u^{\rup*}_{\omega l}(r)$ and $u^\rout_{\omega l}(r) = u^{\rin*}_{\omega l}(r)$.

\subsection{Schwarzschild interior}

We first consider the cavity description of a free field in a classical Schwarzschild interior.  We can find the corresponding $K$ by a Green function construction, but a simpler method is to observe that it is related to internal classical solutions of \eqref{kgeq} by \eqref{respdef}. Working in energy and angular momentum basis then gives
\beq\label{Klo}
K_l(\omega,\rho)=\left[\frac{\partial_r \bar \phi_{\omega l}(r)}{r^2\bar \phi_{\omega l}(r)}\right]_{r=\rho}
\eeq
where here $\bar \phi_{\omega l}$ is the classical wavefunction determined by the internal boundary conditions.  In the case of a Schwarzschild interior, these correspond to pure absorption at the horizon, and so $\bar \phi_{\omega l}(r) = u^{\rin}_{\omega l}(r)/r$.  This fully specifies the cavity response function $K^{\rm Sch}$ corresponding to a classical Schwarzschild interior.

\subsection{Near-horizon reflection}\label{epsmod}

Another boundary condition that has been considered as a simple model imposes a
reflection condition outside the horizon. For example, imposing complete
reflection just outside the horizon has been explored as a model of exotic
compact objects or horizon-scale corrections, yielding late-time gravitational
wave ``echoes''~\cite{Cardoso:2016rao,Abedi:2016hgu,Cardoso:2017cqb,Cardoso:2019rvt,Mark:2017dnq,Price:2017cjr}.  We consider a general such boundary condition including the possibility of partial reflection\cite{FrGi}, at coordinate $r_{*\epsilon}$ corresponding to $r=(1+\epsilon)R$.  This can be imposed by requiring the interior solutions to be of the form
\beq\label{epsbc}
\phi^\epsilon_{\omega l}(r) \propto \phi^\rin_{\omega l}(r) +R^\epsilon_{\omega l}  \phi^\rout_{\omega l}(r)\ ,
\eeq
up to overall normalization,
with an internal reflection coefficient\footnote{Beware the small change of notation from \cite{FrGi}.} $R^\epsilon_{\omega l}$; for small $\epsilon$ we expect this to have a large phase describing propagation to and from the near-horizon boundary, which can be explicitly parameterized as $R^\epsilon_{\omega l} =e^{-2i\omega r_{*\epsilon}}  \hat R_{\omega l}$.  One can show\cite{FrGi} that if the solutions are normalized to unit ingoing amplitude at $r=\infty$ as in \eqref{updown},
\beq\label{epsre}
\phi^\epsilon_{\omega l}= \phi^{\rin}_{\omega l} +  \frac{ R^\epsilon_{\omega l} |T_{\omega l}|^2}{1+{R^\epsilon_{\omega l}} R_{\omega l}^*} \phi^{\rup}_{\omega l}\ ,
\eeq
where $T_{\omega l}$ is the transmission coefficient that complements $R_{\omega l}$ for the potential \eqref{rwpot}.  This form exhibits multiple reflections in the near horizon region, creating echoes, as is further described in \cite{FrGi}.  
Setting $\bar \phi_{\omega l}(r) = \phi^\epsilon_{\omega l}(r)$ in \eqref{Klo} gives the response function $K^\epsilon$ corresponding to this model.

\subsection{Interactions in the BH atmosphere}\label{atmosmod}

An alternative to near-horizon reflection is motivated, for example, by the idea that there could be important corrections in the ``atmosphere" region at  $r=R_a\sim R$, as with non-violent unitarization\cite{Giddings:2017mym,UCNVU}, rather than at scale finely tuned to be microscopically close to the horizon.  
A very simple model assumes  reflecting boundary conditions at such a radius.
 These boundary conditions can be parameterized in the same form \eqref{epsbc}, but it is more natural to parameterize them\cite{FrGi} in the form \eqref{epsre}
\beq\label{ratmos}
\phi^{in}_{\omega l} + \Delta R_{\omega l} \phi^{up}_{\omega l}\ ,
\eeq
with the additional parameterization 
\beq
\Delta R_{\omega l}=  R_{\omega l}(R_a)\, \frac{u_{\omega l}^{\rm in}(R_a)}{ u_{\omega l}^{\rm up}(R_a)}\ ,
\eeq
defining a response function one might call $K^a$ in terms of an effective reflection coefficient $R_{\omega l}(R_a)$.  For example, $R_{\omega l}(R_a)=-1$ corresponds to Dirichlet boundary conditions at $R_a$, and thus total reflection there.

\section{Applications of cavity effective descriptions}

The cavity description of the dynamics can be connected to quantities directly relevant for observation of gravitational wave signatures.  

\subsection{Connection to scattering description}\label{ScattC}

Modifications of classical BH dynamics are expected to alter the scattering of gravitational waves from a BH.  Ref.~\cite{FrGi} introduced a general parameterization of these signatures, in terms of changes of the scattering amplitudes.  In the present context, focusing on ``elastic" interactions (no change in frequency) for the illustrative case of scalars, we can parameterize the general case in terms of the radial wavefunction as 
\beq\label{scattT}
\phi^{\rsc}_{\omega l}  =\phi^\rin_{\omega l} + \Delta \calt_l(\omega)  \phi^\rup_{\omega l} = \phi^\rdown_{\omega l} + \tilde R_{\omega l} \phi^\rup_{\omega l}\ ,
\eeq
with $\Delta \calt_l(\omega)$ denoting the change in the scattering amplitude from that of a BH.  We have also introduced the modified reflection coefficients,
\beq
\tilde R_{\omega l} =R_{\omega l} +\Delta \calt_l(\omega)\ .
\eeq
Each of the preceding examples defines corresponding $\Delta \calt_l(\omega)$'s.
 Combined with \eqref{schbas}, specification of the quantities  $\Delta \calt_l(\omega)$ or $\tilde R_{\omega l}$ leads to definite predictions of the modification of outgoing signals.

This ``S-matrix"-like description is directly related to the action-based description of the cavity effective theory, as the preceding section shows.  To make this connection more explicit, the response function for a given $ \Delta \calt_l(\omega)$ is determined by \eqref{Klo},
\beq\label{TtoK}
K_l(\omega,\rho) = \frac{1}{\rho^2}  \frac{\partial_r \left[\phi^\rin_{\omega l}  + \Delta \calt_l(\omega) \phi^\rup_{\omega l}\right] }{\phi^\rin_{\omega l}  + \Delta \calt_l(\omega) \phi^\rup_{\omega l} }{}_{\Big\vert_{r=\rho}}\ .
\eeq
For example, the cases of \eqref{epsre}, \eqref{ratmos} specify  their associated $ \Delta \calt_l(\omega)$'s.  

Conversely, a given response function determines a corresponding $\Delta \calt_l(\omega)$.  Explicitly, inverting \eqref{TtoK} gives
\beq\label{KtoT}
 \Delta \calt_l(\omega) = -\frac{\phi^\rin_{\omega l}(\rho) \rho^2 K_l(\omega,\rho) - \partial_r \phi^\rin_{\omega l}(\rho)}{ \phi^\rup_{\omega l}(\rho) \rho^2 K_l(\omega,\rho) - \partial_r \phi^\rup_{\omega l}(\rho)}\ .
 \eeq
 Note  that the expression on the right side of \eqref{KtoT} must be $\rho$ independent, connecting to $K$'s running with scale.
 
 The expression \eqref{KtoT} provides a direct link from the action of the cavity effective description to outgoing signals.  A given internal dynamics yields a corresponding response function $K$, and that $K$ determines a correction to the outgoing signal, given a specified ingoing wave profile, through \eqref{KtoT}, \eqref{scattT}, and \eqref{schbas}.

\subsection{Absorption and contributions to binary dephasing}\label{AbsS}

Gravitational waveforms for inspiraling binaries have been computed
systematically in a post-Newtonian (PN) expansion, organized in powers of an
effective orbital velocity.\footnote{See, {\it e.g.}, the reviews
\cite{Blanchet:2013haa,Blanchet:2024pn,Foffa:2013qca,Levi:2018nxp,Porto:2016pyg},
and references therein.} Corrections due to gravitational-wave interaction
with, and absorption by, the individual compact objects also enter these
waveform calculations~\cite{Poisson:1995ef,Alvi:2001mx,Poisson:2004cw,
Goldberger:2005cd,Chatziioannou:2016ezg}. These waveforms are therefore
potentially sensitive to modifications of the objects' response to external
gravitational fields, such as those arising from new physics. Even a small
change in the rate at which the binary loses energy can in principle produce a
significant accumulated change in the signal phase, due to the large number of
cycles in a typical inspiral. Our goal is to extend calculations of such
effects performed in the extreme-mass-ratio case, where the central BH is
treated as stationary~\cite{DaBo,Datta:2019epe}\cite{FrGi}, to comparable-mass systems.   A locally formulated description, like the one we have begun to develop in this paper, provides a way to extend calculation of new physics effects to such general binaries, and to describe possible corrections to GW templates.

Work in this direction was recently carried out in Ref.~\cite{CZI},
which extends earlier studies of possible tidal deformability, or nonzero Love
numbers~\cite{Flanagan:2007ix,Hinderer:2007mb,Binnington:2009bb,
Damour:2009vw,Cardoso:2017cfl,Cardoso:2019rvt,Chia:2023tle}, to include the effects of
tidal heating. The latter terminology arises from a description where gravitational energy from the orbit is absorbed into a material body due to dissipation in its tidal interactions.  A more general perspective, perhaps most pertinent to the case of black holes, is that the relevant effect is simply absorption of the energy of time-dependent gravitational fields -- GWs -- into the object, regardless of a picture involving tides.

The starting point for describing the signal effect of such GW absorption is energy balance; if $\cale$ is the orbital binding energy, energy conservation says it decreases due to mass accretion, $\dot M=\dot M_1+\dot M_2$ of the individual objects, and energy flux $\calf_\infty$ radiated to infinity,
\beq\label{eloss}
\dot \cale = -\dot M - \calf_\infty\ 
\eeq
where dot denotes derivative with respect to the time at infinity (so, Schwarzschild time in the case of a single BH).
This then feeds back into the calculation of the orbital phase and amplitude, which can then be expressed as functions of the PN velocity $v$.

\begin{figure}[h]
 	\begin{center}
 		\includegraphics[width=0.45\textwidth]{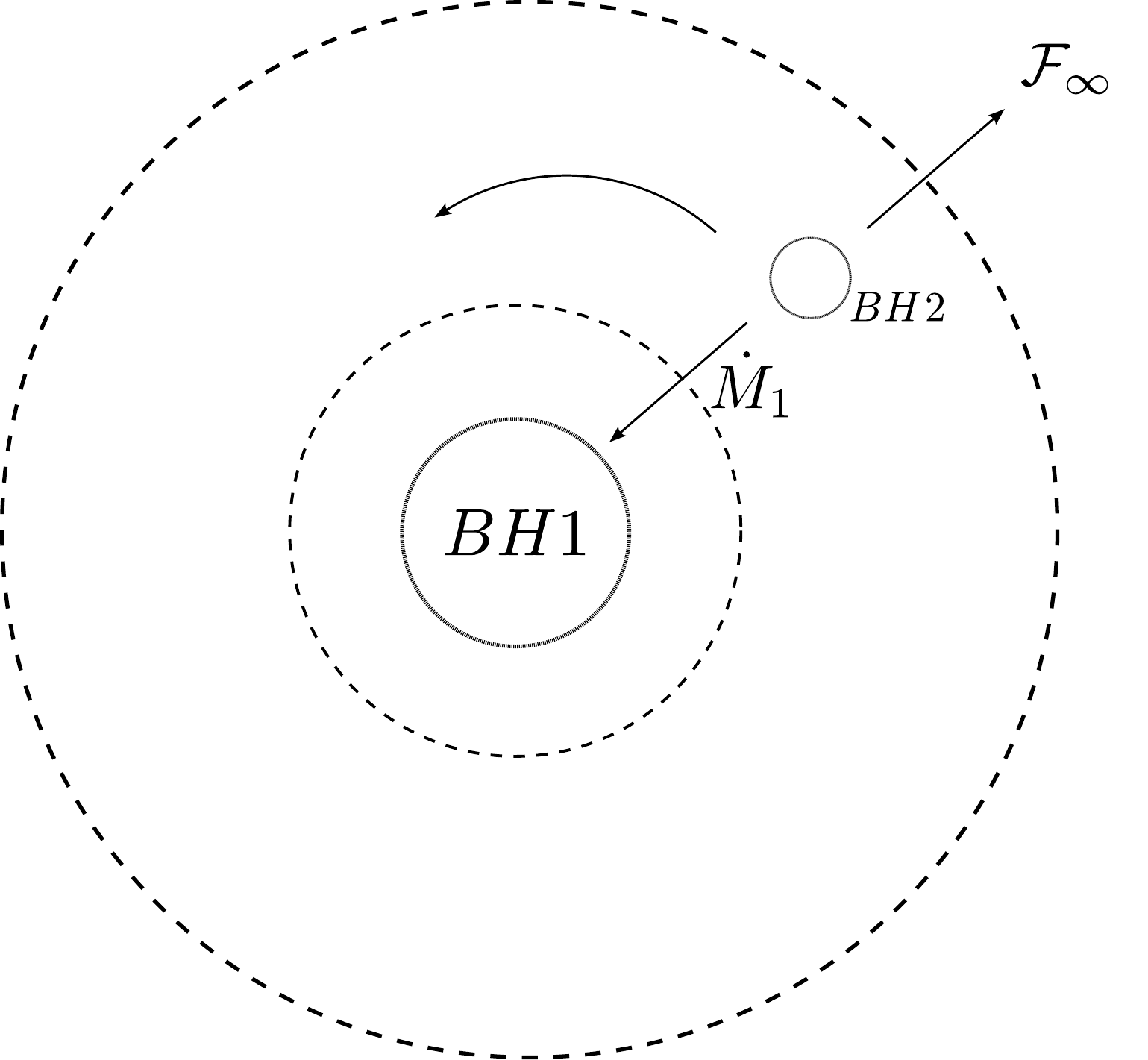} 
 		\caption{Schematic illustration of the energy flux calculations.  A moving source, here BH2, creates radiation, which is either radiated to infinity, or absorbed into the central object, here BH1, increasing its mass. }
 		\label{fig:flux}
 	\end{center}
 \end{figure}

The power loss \eqref{eloss} clearly depends on the interactions of the individual bodies with their surrounding gravitational fields, which we have argued is well-described by their response functions.  Moreover, due to the relatively low velocities, we expect that the effects are dominated by the lower angular momentum components of the response functions.  For example, in the simple case of a circular orbit about a central body,  orbital frequency is related to the mass and orbital radius as
\beq\label{orbf}
\omega_0=\frac{M^{1/2}}{r_0^{3/2}}\ .
\eeq
The orbital frequency determines the emission frequencies.  Eq.~\eqref{orbf} can be rewritten $\omega_0 r_0 = \sqrt{M/r_0}$, so for an orbit in the inspiral phase, where $r_0$ is significantly larger than $R=2M$, $\omega_0 r_0$ is small.  The relative smallness of the emitted frequency is of course directly related to the smallness of $v$, and is correspondingly used as a small parameter in which to expand.

\begin{figure}[t!]
 	\begin{center}
 		\includegraphics[width=0.85\textwidth]{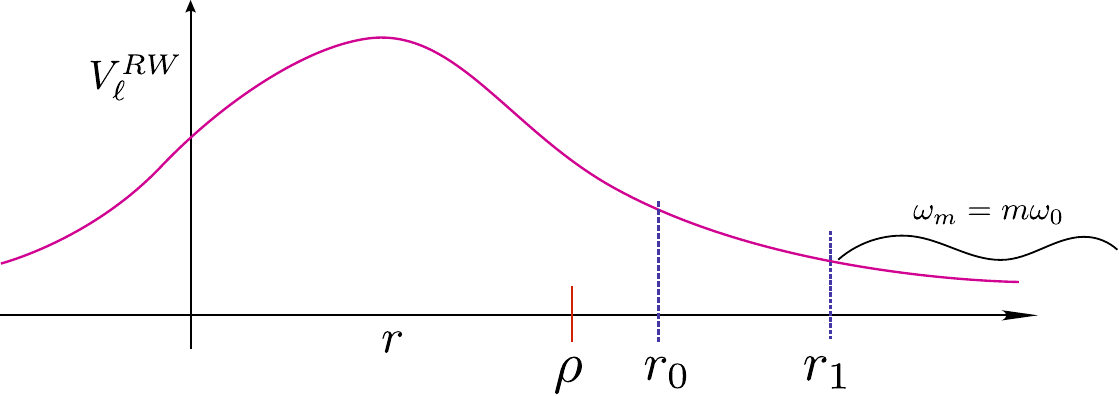} 
 		\caption{Schematic of different scales, associated with emission of radiation governed by effective radial potential $V_l^{RW}$ of 
		\eqref{rwpot}.  The orbital radius $r_0$ lies in a forbidden region of the potential.  Classical wave propagation begins at a larger-radius turning point $r_1$.  We may take the cavity boundary used for matching to lie at a smaller radius $\rho$. }
 		\label{fig:TP}
 	\end{center}
 \end{figure}

We will illustrate the role of the response functions in calculating the power loss in the simple example of scalar radiation; GW radiation is no more complicated conceptually, but has additional calculational complexity that we defer to later work.  In fact, we will find the expected parametric dependence for GW perturbations -- modulo $\calo(1)$ factors -- from the $l=2$ scalar results.

In a gravitational binary, each object serves as a source of radiation that can be emitted to infinity, absorbed by the other object, or reabsorbed; see Fig.~\ref{fig:flux}.  To study the role of the cavity approach in describing this emission and absorption, consider the problem of a general classical source $J(x)$ in the background of a single object.  The resulting classical solution $\phi_J$ is found from the classical equation of motion derived from the cavity description \eqref{zwbdy}.  The stress tensor for $\phi_J$ then describes the energy flux to infinity, and into the object.

In the case of a free (quadratic) boundary action, the solution $\phi_J$ is determined from the corresponding Green function, which has already been found from the scattering description in \cite{FrGi}, and so can be rewritten in terms of the response function via \eqref{KtoT}.  

The power loss \eqref{eloss} can then be calculated from the energy flux to infinity or into the object, calculated by integrating the stress tensor for $\phi_J$ over the corresponding surfaces.  Adapting the calculation of \cite{FrGi} (appendix A) to wavefunctions of the form \eqref{scattT} 
gives an expression
\beq\label{ploss}
-\langle\dot \cale\rangle_T = \sum_{lm} \int_0^\infty d\omega d\omega' \Delta_T(\omega-\omega') 2\omega^2 \left\{ \big |Z^{{\bar{\rsc}}}_\olm[J]\big|^2 + |\tilde T_{\omega l}|^2 \big |Z^{down}_\olm[J]\big|^2\right\}\ ,
\eeq
where the first and second terms correspond to $\calf_\infty$ and $\dot M$, respectively.  Here we have averaged over a time interval $T$, and in particular define
\beq
\Delta_T(\omega-\omega') = \frac{1}{T} \int_{-T/2}^{T/2} dt e^{-i(\omega -\omega') t}\ ,
\eeq
which is for example $\propto\delta(\omega-\omega')$ in the large $T$ limit. The magnitude of the two energy-loss terms is governed by the overlap integrals with the source
\beq
Z^{down}_\olm[J] = \int dV_4 \frac{\phi^{down*}_{\omega lm}(x)}{4\pi i\omega} J(x)\ , \ Z^{{\bar{sc}}}_\olm[J] =  \int dV_4 \frac{\phi^{\bar{\rsc}*}_{\omega lm}(x)}{4\pi i\omega} J(x)\ ,
\eeq
defined in terms of the full wavefunctions \eqref{schbas}, \eqref{scattT} using also the definition
\beq
u^{\bar{\rsc}}_{\omega l}(r) = u^{sc*}_{\omega l}(r)\ ,
\eeq
analogous to the relation $u^\rout_{\omega l}(r) = u^{\rin*}_{\omega l}(r)$.
Finally $\tilde T_{\omega l}$ is the transmission coefficient complementary to the reflection coefficient $\tilde R_{\omega l}$ of \eqref{scattT}, with in particular $|\tilde R_{\omega l}|^2 + |\tilde T_{\omega l}|^2=1$.  The change in power loss \eqref{ploss} can also be expressed in terms of the deviation $\Delta \calt_l(\omega)$ from classical BH behavior, as described in \cite{FrGi}.  In the case of periodic orbits, producing discrete radiation frequencies $\omega_n$, \eqref{ploss} becomes 
\beq\label{plossp}
-\langle\dot \cale\rangle = \sum_{nlm}  2\omega_n^2 \left\{ \big |Z^{{\bar{\rsc}}}_{nlm}[J]\big|^2 + |\tilde T_{\omega_n l}|^2 \big |Z^{down}_{nlm}[J]\big|^2\right\}\ .
\eeq

As part of understanding the formulas \eqref{ploss}, \eqref{plossp}, note that orbits may typically be in the ``near-zone" region under the potential barrier for radiation propagation.  For example consider a circular orbit of a pointlike object with scalar charge $q$, with orbital radius $r_0$ and frequency $\omega_0$ related by \eqref{orbf}, taken to be in the plane $\theta=\pi/2$.  This gives\cite{FrGi}
\beq
J(x)= \frac{q}{r_0^2} \frac{d\tau}{dt}\Big\vert_{r_0} \delta(r-r_0) \delta(\theta-\pi/2) \delta(\phi-\omega_0 t)\ ,
\eeq
and in terms of the relevant radial wavefunctions,
\beq\label{Zcirc}
Z_{\omega l m}[J] = {q}\frac{d\tau}{dt}\Big\vert_{r_0} \frac{u^*_{\omega l}(r_0)}{2i\omega r_0} Y^*_{lm}(\pi/2,0) \delta(\omega-m \omega_0)\,.
\eeq
The maximal frequency for given $l$ is therefore $l\omega_0$.  Assuming any modifications to the potential $V_l(r)$ are at $r<r_0$, and in the case of massless radiation, the turning point for the radiation is at $r_1\sim\sqrt{l(l+1)}/\omega$, and so has a minimum value of $\sim 1/\omega_0 \gg r_0$ (see Fig.~\ref{fig:TP}). In the case of the flux to infinity, the first term in \eqref{plossp} then has an exponentially small tunneling factor $T(r_0,r_1)$, arising from propagation under the barrier, due to the wavefunction factor from \eqref{Zcirc}.  Likewise, the solution $u^{\bar\rsc}$ {\it grows} exponentially from unit magnitude towards smaller $r$ under the barrier; the corresponding wavefunction factor from \eqref{Zcirc} partially cancels the tunneling factor $\tilde T_{\omega l}$ to give a net exponentially small tunneling factor $T(r_0,r_1')$ to the analogous internal turning point $r_1'$.  

Due to this near-zone behavior, the dominant modification to the power drain from orbital energy \eqref{eloss} arising from modifications to BH behavior arises from the change to $\dot M$.  The power loss, and in particular $\dot M$, may also be simply described in terms of the cavity effective theory and its response function.

Specifically, the integrated energy flux across the cavity boundary in general is
\beq\label{MdotF}
\dot M  = \oint_\surf d\surf\frac{d\tau}{dt}\, n^\mu T_{\mu t}  =   \oint_\surf d\surf\frac{d\tau}{dt}\, \nabla_n \phi \partial_t\phi\ 
\eeq
where we have included the redshift factor $d\tau/dt$ to proper time at the cavity boundary.
The cavity boundary condition \eqref{cbc} then gives
\beq\label{Mdcbc}
\dot M =  -\oint_\surf d\surf\frac{d\tau}{dt}\,  \frac{\delta S_\partial}{\delta \varphi} \partial_t\phi \ ,
\eeq
which in the non-interacting case can be written in terms of $K$ using \eqref{bdfixni},
\beq\label{Mdni}
\dot M =   \oint_\surf d\surf\frac{d\tau}{dt}\,  \partial_t\phi \oint_\tube d\tube' K(x,x') \phi(x')\ .
\eeq

Again considering the simplification of spherical symmetry, we expand the boundary values of $\phi$ and $\nabla_n\phi$ as
\bea\label{bdexp}
\varphi(t,\Omega) &=& \int \frac{d\omega}{2\pi}\sum_{lm} \varphi_{\omega l m} e^{-i\omega t} Y_{lm}(\Omega) + cc\cr 
\nabla_n\phi_{\vert\tube} &=& \int \frac{d\omega}{2\pi}\sum_{lm} (\nabla_n\phi)_{\omega l m} e^{-i\omega t} Y_{lm}(\Omega) + cc\ .
\eea
In this case, \eqref{bdfixni} simplifies to
\beq\label{bcom}
(\nabla_n\phi)_{\omega l m} = \rho^2 \sqrt{f(\rho)} K_l(\omega, \rho)\varphi_{\omega l m}\ .
\eeq
The time-$T$ average of \eqref{Mdni} then becomes
\beq\label{mdotav}
\langle \dot M\rangle_T = -2\rho^4 f(\rho)\int \frac{d\omega}{2\pi} \frac{d\omega'}{2\pi} \omega \Delta_T(\omega-\omega') \sum_{lm} |\varphi_{\omega l m}|^2 {\rm Im} K_l(\omega,\rho)\ .
\eeq

This derivation has a close parallel with calculations described in the worldline EFT approach\cite{Goldberger:2005cd,GLR,CZI}, using the quadrupole couplings \eqref{sublead}.  The quantities $\nabla_n\phi_{\vert\tube}$, fixed by \eqref{cbc}, \eqref{bdfixni}, or \eqref{bcom}, are analogs to the  moments $Q_{ab}$ of \eqref{sublead}, as can also be seen via \eqref{afix}, and analogously couple to the boundary field $\varphi$ via the cavity action $S_\partial$.  Specifically, the worldline EFT calculation of $\langle Q_{ab}\rangle$ in terms of $E$ or $B$ is analogous to \eqref{cbc}, \eqref{bdfixni}, or \eqref{bcom}.  Then, equations \eqref{Mdcbc}, \eqref{Mdni} are analogous to expressions for the mass gain in \cite{GLR,CZI}.  

However, eqs.~\eqref{Mdcbc}, \eqref{Mdni}  extend those worldline EFT expressions since they contain couplings to all multipoles of the fields, \eqref{bdexp}, and not just the quadrupole coupling of \eqref{sublead}.  For a source corresponding to an object orbiting at radii $\gg \rho$, we expect suppression of the contributions of higher multipoles, with suppression factors $\sim \rho^l$ arising from the contribution of the radial wavefunctions $u_{\omega l m}(\rho)$ to $\varphi_{\omega lm}$.

If we consider a low frequency source, {\it e.g.} corresponding to the relatively low orbital frequency \eqref{orbf}, we can approximate $K_l(\omega,\rho)$ in \eqref{bcom} by its low-frequency expansion,
\beq
K_l(\omega,\rho) = \sum_n K^n_l(\rho) \omega^n \simeq  K^0_l(\rho) +  K^1_l(\rho)\omega\ .
\eeq
The zero-frequency quantities $K^0_l(\rho)$ determine the Love numbers, through the formula
\beq\label{KtoL}
\hat \lambda_l = -\frac{\rho^2 K_l^0(\rho) P_l(x_\rho) -\partial_\rho P_l(x_\rho)}{\rho^2 K_l^0(\rho) Q_l(x_\rho) -\partial_\rho Q_l(x_\rho)}\
\eeq
where $P_l$ and $Q_l$ are Legendre functions and $x_\rho=2\rho/R-1$.  These are derived in Appendix~\ref{LApp}; the simple expression \eqref{KtoL} gives them in a non-standard normalization, and conversion to more standard normalization also appears there.
These are real, and vanish for a classical BH background.   The leading contribution to $\dot M$ in \eqref{mdotav} thus arises from the quantities $K^1_l(\rho)$.

For scalars there can be a leading monopole contribution to the mass growth, but to model the GW case with leading quadrupole absorption we can focus on $l=2$.  Approximating \eqref{bcom} by
\beq\label{approxresp}
(\nabla_n\phi)_{\omega 2m} \simeq \rho^2 \sqrt{f(\rho)} \left[K^0_2(\rho) +  K^1_2(\rho)\omega\right] \varphi_{\omega 2 m}
\eeq
and comparing with the worldline EFT expression \cite{Goldberger:2020wbx}\cite{CZI} (with $G=1$)
\beq\label{WLQ}
\langle Q^E_{ab}\rangle = M^5\left[c_0 E_{ab} + M H_\omega \frac{dE_{ab}}{d\tau}\right]
\eeq
shows that the coefficient $K_2^1$ gives a scalar analog of the ``tidal heating coefficient" defined there\footnote{
The precise comparison of \eqref{approxresp} with \eqref{WLQ} is obtained by converting the response function to asymptotic source and response data,
\(\phi_{\omega2m}\sim \mathcal E_{\omega2m}r^2+\mathcal Q_{\omega2m}r^{-3}\).
Then the scalar analog is
\(
\mathcal Q_{\omega2m}
=
M^5\left[
 \lambda_2
-i
M H_2^\phi \omega+\cdots
\right]\mathcal E_{\omega2m},
\)
so that \( \lambda_2 \) is the scalar Love coefficient, while
\(H_2^\phi \) is the dissipative coefficient extracted from the properly
normalized \(O(\omega)\) part of \(K_2\). In other words, $H_2^\phi=
\frac{i}{M^6}
\left.
\frac{d}{d\omega}
\left(
\frac{\mathcal Q_{\omega 2m}}{\mathcal E_{\omega 2m}}
\right)
\right|_{\omega=0}.$
},
\beq
 H_\omega \leftrightarrow -H_2^\phi = -\frac{i\rho^8}{5M^6} K^1_2 \ .
\eeq
The Love number,  $c_0$ in \eqref{WLQ}, is determined as above. 

These expressions can be illustrated in the case of a classical BH, where the response function is specified in terms of known Heun functions via eq.~\eqref{Klo}.  The result has low-frequency expansion
\begin{align}
K_2^{\rm BH}(\omega,\rho)
&=
\frac{
6(2\rho-R)
}{
\rho^2\left(6\rho^2-6R\rho+R^2\right)
}
-
\frac{
iR^6\omega
}{
\rho^3(\rho-R)
\left(6\rho^2-6R\rho+R^2\right)^2
}
+\mathcal O(\omega^2) .
\end{align}
The corresponding Love numbers vanish, as noted above.
Taking the cavity boundary at $\rho\gg R$ gives
\beq
K_2^1 \approx -i\frac{R^6}{36\rho^8} + \calo\left(\frac{R^7}{\rho^9}\right)\ ,
\eeq
and so $H_2^\phi=16/45$, to be compared to the expression\cite{Goldberger:2005cd,CZI} $H_\omega=16/45$ for GWs.

Working in the approximation where we only consider the quadrupole coupling and frequencies are low, and also considering discrete frequencies $\omega=\omega_n$ as for periodic orbits, the formula for the average power loss \eqref{mdotav} becomes
\beq\label{mdapprox}
\langle \dot M\rangle_T = 10\frac{M^6 f(\rho)}{\rho^4}  H_2^\phi \sum_{nm} \omega_n^2 |\varphi_{n2m}|^2\ .
\eeq
To calculate actual rates, one needs the scalar field profile, which depends on the orbit.  For example, for a point source with charge $q$ in a circular orbit, the corresponding field profiles have been computed using Green functions in \cite{FrGi}.  The formula \eqref{mdapprox} is a direct analog of the corresponding worldline-EFT based formula for GWs~\cite{Goldberger:2005cd} considered in \cite{CZI} (keeping only the leading electric contribution),
\beq\label{gwabs}
\dot M = M^6 H_\omega \dot E^{ij} \dot E_{ij}\ .
\eeq
Beginning with this formula, and calculating the tidal moments for a binary orbit as in \cite{TaPo,CPY1,CPY2}, one may find the leading PN contribution to $\dot M$ for each binary member, in terms of the parameters $M_1$, $M_2$, and PN velocity $v$ of the binary, and the absorption coefficient $H_\omega$.  

The energy absorption \eqref{mdapprox} or \eqref{gwabs} then contributes to the decay of the orbital energy, via the energy conservation equation, \eqref{eloss}.   
For example, for nearly-circular orbits the GW phase is calculated in a PN expansion using Kepler's law $\dot \Phi = v^3/M$, with $\Phi$ denoting the orbital phase, with $M=M_1+M_2$.
The leading contribution from the mass gain gives
\beq
\dot v \simeq -\left(\frac{\partial \cale}{\partial v}\right)^{-1}\left[\calf_\infty + \dot M_1+ \dot M_2\right]\ .
\eeq
Combining this with the expression for the phase,
\beq
\Phi(v)= \Phi_0 + \int dv \frac{v^3}{M}\frac{1}{\dot v}
\eeq
then gives the corresponding correction to the phase $\Phi(v)$ as a function of $v$, or of frequency, which begins at 4PN order.

This provides a direct route from the cavity effective theory and its response function to gravitational template corrections.  We have not explicitly performed the last step of this route in the scalar model, by calculating the scalar profile for binary orbits, since the more immediate goal is the analogous treatment of gravitational waves.  

Moreover, if classical BH behavior is modified by some other, {\it e.g.} quantum, effects, that can be described through a corresponding modification to $S_\partial$ or $K$, and this modification converted into a corresponding correction to GW templates.

To illustrate such modifications, we can consider the simple models described in section \ref{csec}.  

For the case of near-horizon reflection of Sec.~\ref{epsmod}, $K_l(\omega,\rho)$ is given by \eqref{TtoK}, with $\Delta \calt_l(\omega)$ given by the coefficient of $\phi^\rup_{\omega l}$ in the second term of \eqref{epsre}.  In the near-horizon limit where $\epsilon$ is small, this contains terms with a large phase, {\it e.g.} $\Delta   \calt_l(\omega) \propto e^{-2i\omega r_{*\epsilon}}$, as indicated there, and likewise for terms in the denominator.  This corresponds to dynamics on long timescales $\sim r_{*\epsilon}\sim -R\log\epsilon$.  While this produces new long-term effects, such as echoes, these terms are expected to average to zero (be ``out of phase") when integrated against signals of much shorter duration, {\it e.g.} of orbital timescales $\sim 1/\omega_0$.  The expected result is that these terms do not change the energy loss rate over orbital periods, which is then the same as for a classical BH.  (If the inspiral timescale exceeds the timescale $\sim r_{*\epsilon}$, we expect the corresponding ``echo energy" to then radiate to infinity, contributing to the direct GW signal but not additionally to change of phase.)

For the case of scattering from the BH atmosphere, as in Sec.~\ref{atmosmod}, complete reflection arising from $R_{\omega l}(R_a)=-1$ means that there is {\it no} mass accretion,  due to the Dirichlet boundary conditions and therefore, by \eqref{MdotF}, vanishing $\dot M$.  One may alternately choose Neumann boundary conditions $\nabla_n\phi_{\vert R_a}=0$, with the same effect.   Indeed, the latter, Neumann, boundary conditions correspond to $K$ vanishing at the atmosphere scale,
$K_l(\omega,R_a)$=0.  Intermediate values, $|R_{\omega l}(R_a)|<1$, will produce intermediate, reduced, absorption from the BH value.
However, the flux $\calf_\infty$ to infinity is also modified.  The net effect is a correction to the energy loss rate that can exceed $\dot M$ by around two orders of magnitude\cite{FrGi}.  This enhancement can be understood as a result of less tunneling suppression to  the boundary at $R_a$ as compared to reaching the horizon of the BH, and raises interest in study of models of this kind.
These models also produce nonvanishing Love numbers.  GW template corrections from Love numbers and initial exploration of resulting observational bounds appear in \cite{Chia:2023tle}.

\section{Further directions}

This paper has focused on the conceptual problem of relating detailed, ``microscopic" descriptions of dynamics to longer-distance effective descriptions, and thus to observational signatures.  For these conceptual purposes, we have focused on the simplest cases: spin zero fields, nonspinning black holes, and primarily freely propagating fields.  This is preparation for extension of the basic lessons to more observationally relevant cases, specifically tensor perturbations, spinning BHs, and possible modifications to classical BH dynamics. We comment further on these here, and anticipate addressing them in further work.


Analogous treatment of spin two GW emission follows a parallel path to the scalar case, with the main technical complications being treating polarizations and gauge invariance.  Specifically, we expect there to be a decomposition of the functional integral, as in \eqref{genfeq}, and a resulting boundary action analogous to \eqref{bfix} or \eqref{afix}.  In the approximation where the gravitational perturbations are treated as freely propagating on a background, one expects  corresponding response functions, $K$, as in \eqref{sbdy}.  These perturbations may be treated about the original BH background, or about a modified (regulated) ``point particle" background, as in Sec.~\ref{PPsec}, making contact with the worldline EFT approach.  And, we likewise expect a cavity effective description, as outlined in Sec.~\ref{CED}, to be useful in describing long-wavelength dynamics, as for inspiralling binaries.  This cavity description should be related to the analogous scattering description for GWs described in \cite{FrGi}, as in Sec.~\ref{ScattC}.  Finally, the cavity description should also give an approach to calculating contributions to the gravitational wave signal, both direct, and also through phase corrections due to GW absorption into the individual bodies, also connecting to~\cite{FrGi}, as in Sec.~\ref{AbsS}.

Spinning BHs are expected to require more technology, and in particular a frame orientation associated with the spin, but are expected to likewise have cavity descriptions like those of this paper.

A key goal is to describe possible modifications of classical BH dynamics, specifically those associated with the ultimate need for BHs to evolve in a way that is consistent with quantum mechanics. There are different proposals of this nature, ranging from classical modifications to BH structure, to fuzzballs\cite{Mathur:2005zp,Skenderis:2008qn,Bena:2007kg,Bena:2022rna}, to wormhole/baby universe dynamics\cite{Coleman:1988cy,Giddings:1988cx,PSSY,AHMST,MaMa,Almheiri:2020cfm}, to nonviolent unitarization\cite{Giddings:2017mym,UCNVU}.  These generically suggest modifications of the way a BH interacts with its surrounding gravitational fields.  We expect that if such modifications  alter GW signals from classical BH expectations, a route to describing these alterations is via an effective description, and in particular via modifications to the cavity descriptions we have given in this paper.
Thus, we would like to understand and parameterize such effects in a description in terms of the boundary action $S_\partial$, or response function $K$.  We have briefly studied very simple models for such modification, but expect more realistic models to have richer physical behavior.


In particular, nonviolent unitarization\cite{Giddings:2017mym,UCNVU} provides a class of models with unitarizing dynamics to restore quantum mechanical BH evolution.
We expect to be able to extend the approach of this paper to treating more realistic models for the dynamics of NVU.  
The simple models of section \ref{atmosmod} do describe interactions at atmosphere scales, but we might expect other features to be present.  Specifically, if there are interactions between BH quantum states and atmosphere modes, these are expected to nontrivially alter the action $S[\phi]$, or its generalization to tensor perturbations, in the vicinity of the BH; these effects could then be parameterized as contributions to
the corresponding boundary action $S_\partial$ defined either by \eqref{bfix} or \eqref{afix}, or, in the linear case, to the corresponding response function $K$. In this latter case, we expect the response function to be related to the modified two-point function, generalizing the relations described for the minimal $S[\phi]$.  
And, for example, if these new interactions lead to greater absorption, that would be described by a larger low frequency ${\rm Im} K$.  In future work, we anticipate investigating models for these underlying interactions, and their effects of modifying $S_\partial$ or $K$.  Notice that the absorption described by, for example, nonvanishing ${\rm Im} K$ in effect arises from treating the system as being open.  A more complete accounting of the full unitary dynamics would include the coupling to BH quantum states and their evolution.  We expect that the formalism of this paper can also be extended to one that accounts for these BH states, {\it e.g.} as parameterized in \cite{GiPo}, and their effective interactions.

It also appears useful to extend cavity descriptions to directly incorporate the position and motion of the individual objects.  The description we have developed is based on a cavity that is localized on an object which we may by choice of coordinates take to be at rest.  With two such objects, we would like to also parameterize their relative motion.  We expect that this can be incorporated through dependence of the boundary action $S_\partial$ or response function $K$ on effective collective coordinates parameterizing the location of the cavity, and anticipate further development in this direction.


Finally, the effective descriptions studied in this paper involve several
auxiliary scales, and their scale dependence suggests natural RG
interpretations. First, in deriving a worldline EFT from a theory regulated at a
short-distance scale \(\bar\rho\), individual terms in the effective action
depend on \(\bar\rho\). This dependence is not physical: it is compensated by
the running of Wilson coefficients, so that cavity observables and asymptotic
quantities are independent of $\bar \rho$. In this sense, Love numbers, dissipative
coefficients, and higher-derivative response coefficients can be viewed as
scale-dependent couplings whose RG flow encodes how short-distance physics is
reorganized as the cutoff is changed. Second, the cavity description contains a distinct matching scale, denoted here
by \(\rho\). Varying \(\rho\) changes the finite-radius response function, but not
the outgoing physical solution. The resulting radial flow is therefore an RG-like
evolution in which the response function runs with the cavity radius. In practice this
flow is governed by  radial equations of motion, or equivalently, for the Love numbers, by a
Riccati equation for the logarithmic derivative of the field. (See Appendix \ref{LApp}.) The fixed points
of this radial flow are associated with the independent asymptotic branches,
while physical response data are encoded in RG-invariant combinations that
relate these branches. These two forms of scale dependence are conceptually distinct. The
\(\bar\rho\)-dependence arises from short-distance renormalization of worldline
couplings, whereas the \(\rho\)-dependence is a matching-scale dependence of
the finite-radius cavity kernel. Nevertheless, both forms of scale dependence
are expected to cancel from physical observables. Specifically, quantities such as
\(\Delta\mathcal T\), which relate asymptotic in- and out-states and therefore
have an S-matrix-like character, should be invariant under changes of the
cavity radius after the corresponding kernel is evolved consistently. This
suggests that the formalism developed here may provide a useful setting for
studying gravitational response in terms of RG flow, with Love numbers and
dissipative coefficients appearing as invariant data extracted from
scale-dependent effective descriptions.

\section*{Acknowledgments}
This work was supported in part by the Heising-Simons Foundation under grants \#2021-2819 and \#2024-5307, and by the U.S. Department of Energy, Office of Science, under Award Number {DE-SC}0011702.  We thank Y. Chen, R. Porto, B. Seymour, and T. Venumadhav for useful conversations.

\appendix

\section{Outline of explicit connection to worldline EFT}\label{KtoEFT}

In cavity effective descriptions, the dynamics is characterized in terms of field perturbations and responses at the cavity boundary $\tube$, which are summarized by the boundary effective action $S_\partial$, and, in the linear case, response functions $K$.  In the worldline EFT description, as originally envisioned, the dynamics is described in terms of operators on the worldline.  However, as we have emphasized, the leading worldline action \eqref{ppact} produces solutions with no regular center at which to place these operators.  Moreover, these solutions already incorporate the dynamics -- such as absorption -- that higher order terms in the EFT seek to parameterize.  

One can view these problems as arising from the absence of a cutoff prescription for the worldline EFT.  We can modify the action \eqref{ppact} to incorporate such a cutoff, either by modifying the fundamental behavior of the action, such as in treatments of the Wilson-Polchinski renormalization group~\cite{Polchinski:1983gv}, or by adding additional fields that regulate the short distance behavior, an example being matter producing a star-like solution.  Once one has done so, one can have  a regular origin at which to base EFT analysis.

In the spherically-symmetric case, we parameterized the resulting regulated metric $\bar g_{\mu\nu}$ in \eqref{regmet}.  One then considers perturbations about this, and summarizes the corrections due to BH or other related behavior through an effective action $\Delta S_\partial[\varphi]$ as in \eqref{genfp}.

The action $\Delta S_\partial[\varphi]$ still involves the fields on the boundary $\tube$ of the cavity, rather than at the origin.  However, now that one {\it has} a regular origin, described with metric $\bar g_{\mu\nu}$, one can translate to local operators at this origin.  Specifically, consider the spherically-symmetric case with boundary at $r=\rho$.  Define the angle-dependent unit vectors corresponding to points on $S^2$ as $e^i(\Omega)$, and introduce spatial coordinates $x^i=r e^i(\Omega)$.  The field can then be expanded in a Taylor series, resulting in the $r=\rho$ expression
\bea
\varphi(t,\Omega)=\phi(t,\rho,\Omega) &=& \phi(t,0) + \rho e^i \partial_i \phi(t,0) + \frac{\rho^2}{2} e^ie^j \partial_i \partial_j \phi(t,0) + \cdots\cr
&=& \sum_{p=0}^\infty \frac{\rho^p}{p!} E^P(\Omega) \partial_P\phi(t,0)\ ,
\eea
where $P$ denotes a multi-index, $\{i_1,\cdots,i_p\}$, and $E^P(\Omega)$ is the corresponding product of the $e^i(\Omega)$.  Using this expansion in the boundary action $\Delta S_\partial[\varphi]$ converts the latter  into an expression in terms of local operators at $r=0$, with a derivative expansion.  For example, in the free case \eqref{sbdy}, and using spherical symmetry, $\Delta S_\partial$ can be written in terms of the $\Delta K_l(\omega,\rho)$ like in  \eqref{KSexp}.  The integrals of $\Delta K$ against $\varphi$ at the boundary  $r=\rho$ then yield, {\it e.g.}, 
\beq
\int d\Omega Y_{lm}^*(\Omega) \varphi(t,\Omega) = \sum_{p=0}^\infty \frac{\rho^p}{p!} \left[\int d\Omega Y_{lm}^*(\Omega) E^P(\Omega) \right] \partial_P\phi(t,0)\ .
\eeq
The integrals in the square brackets project onto traceless combinations of the $E^P$'s, with known coefficients~\cite{Thorne:1980ru}.  For a given $l$, the trace terms with contractions among indices in $P$ yield higher derivatives $\partial^2_r$, {\it etc.}  This connects the multipole expansion of the cavity effective theory to the derivative or momentum expansion of the worldline EFT, and concretely re-expresses the boundary action $\Delta S_\partial$ in terms of an expansion in operators quadratic in $\phi$ with increasingly many derivatives.  Given these expressions, and metrics $g_{\mu\nu}$ and $\bar g_{\mu\nu}$, defining $\Delta K$'s, various definite examples may be worked out.  In addition to the examples where $g_{\mu\nu}$ is a BH metric, one may also consider illustrative toy examples where $\bar g_{\mu\nu}$ is the flat metric, and $g_{\mu\nu}$ contains some metric perturbations.

\section{Love numbers}\label{LApp}
\subsection{Calculation of Love numbers}
The stationary tidal response of an object is characterized by its Love numbers; these have been extensively studied in the gravitational context.  For the present simple example of scalar perturbations of non-spinning bodies, a useful starting point is the zero-frequency version of the Klein-Gordon equation \eqref{kgeq}, with $m=0$.  With solutions of the form \eqref{schbas}, we may work with the radial equation \eqref{radeqnt}, or alternately directly with the equation for $\phi_l(r)=\phi_{\omega l}(r)_{\vert\omega=0}$,
\beq
\frac{1}{r^2}\partial_r\!\left[r^2 f(r)\,\partial_r \phi_l\right]
-
\frac{l(l+1)}{r^2}\,\phi_l
= 0\ .
\eeq
The change of variable
\beq
x = \frac{2r}{R} - 1
\eeq
then transforms this to Legendre's equation,
\beq
   [(1-x^2)\partial_x^2-2 x\partial_x +l(l+1)]\phi_l=0\ ,
\eeq
with the horizon at $x=1$, and with solutions that are linear combinations of the standard Legendre functions $P_l(x)$, $Q_l(x)$.  These are related to the zero frequency limit of the in and up wavefunctions as
\beq
\frac{\phi^\rin_{\omega l}}{T_{\omega l}} \rightarrow P_l(x)\quad ,\quad \frac{\phi^\rup_{\omega l} T_{\omega l}}{2 i\omega}\rightarrow Q_l(x)\ ,
\eeq
with $T_{\omega l}$ the transmission amplitude through the Regge-Wheeler potential $V_l^{RW}$, as may be seen by comparing direction of growth under the barrier and normalizations.

A given object will thus produce radial solutions of the general form
\beq
\phi_{ l}(r)\propto P_l(x) + \hat \lambda_l Q_l(x)\ ,
\eeq
determined for example from the response function $K$.  One definition of the Love numbers is as the relative coefficients $\hat \lambda_l$.
These are related to the response function through the zero-frequency version of \eqref{Klo}, which may be solved to give a direct analog to  eq.~\eqref{KtoT} for $\Delta \calt_l(\omega)$, 
\beq
\hat \lambda_l = -\frac{\rho^2 K_l(0,\rho) P_l(x_\rho) -\partial_r P_l(x_\rho)}{\rho^2 K_l(0,\rho) Q_l(x_\rho) -\partial_r Q_l(x_\rho)}\ ,
\eeq
with $x_\rho=2\rho/R-1$.  

A more standard normalization of the Love numbers\cite{Binnington:2009bb} is as the relative coefficients in the asymptotic $r\rightarrow\infty$ behavior,
\beq\label{Lnorm}
\phi_l(r)\rightarrow A_l( r^l + \lambda_l r^{-l-1})\ .
\eeq
The known asymptotics of $P_l$ and $Q_l$,
\beq\label{PQas}
P_l(x)\sim \frac{(2l)!}{2^l (l!)^2} x^l\quad, \quad Q_l(x)\sim \frac{2^l (l!)^2}{(2l+1)!} x^{-l-1}
\eeq
give the conversion
\beq
\lambda_l =  \frac{(l!)^4 }{2(2l+1)!(2l)!} R^{2l+1}\hat \lambda_l\ .
\eeq

For classical BHs, these definitions show the Love numbers vanish, as first observed in \cite{Binnington:2009bb,Damour:2009vw}.  Models with reflection in the atmosphere supply an example with nontrivial Love numbers.  These are easily found, {\it e.g.} from the Dirichlet condition $\phi_l(R_a)=0$, or its generalzations for example with partial reflection:
\beq
\hat \lambda_l = R_{0l}(R_a) \frac{P_l[x(R_a)]}{Q_l[x(R_a)]}\ .
\eeq
The asymptotic behavior \eqref{PQas} shows that for large $R_a$,
\beq
\lambda_l\sim R_{0l}(R_a) R_a^{2l+1}\ .  
\eeq
One can also calculate the Love numbers for near-horizon reflection, but in that case the $\omega\rightarrow0$ limit has a more subtle interpretation in orbital dynamics, as discussed in the main text.

\subsection{Renormalization group perspective}

The cavity description also supplies a renormalization group perspective on the calculation of the Love numbers. 
Rewrite \eqref{Klo} as
\begin{align}
\frac{\phi_l'(r)}{\phi_l(r)}=r^2K_l(r,\omega).
\end{align}
The radial equation then implies the Riccati flow
\begin{align}
K_l'
=
\frac{l(l+1)}{r^4 f}
-
\frac{\omega^2}{r^2 f^2}
-
r^2K_l^2
-
\left(
\frac{4}{r}+\frac{f'}{f}
\right)K_l .
\label{eq:riccati_K}
\end{align}
At finite frequency the near-horizon flow has two branches,
\begin{align}
K_l(r,\omega)
\simeq
\pm \frac{i\omega}{R(r-R)},
\end{align}
corresponding to \(e^{\pm i\omega r_*}\).
The black-hole boundary condition selects the ingoing branch.
In the static problem this becomes the requirement that the field
itself be regular at the horizon, selecting \(P_l(x)\) and
excluding the logarithmically singular \(Q_l(x)\).

At large radius, define the dimensionless running coupling
\begin{align}
g_l(r) = r^3 K_l(r,\omega),
\end{align}
whose flow equation in the static limit is
\begin{align}
r\,\frac{dg_l}{dr}
= -(g_l - l)(g_l + l + 1)
+ \mathcal{O}(R/r).
\label{eq:g_flow}
\end{align}
The static flow has two asymptotic fixed points,
\(g_l = l\) and \(g_l = -(l+1)\), corresponding to
\(r^l\) and \(r^{-l-1}\).
For solutions with nonzero source amplitude, the trajectory
approaches the source fixed point.
The Love number is proportional to 
the coefficient of the subleading response
deformation in this approach:
\begin{align}
g_l(r) = l + C_l\,r^{-(2l+1)} + \cdots,
\qquad
C_l \propto \lambda_l,
\end{align}
where the large-radius field expansion is given in \eqref{Lnorm}.
For a Schwarzschild black hole, horizon regularity sets  \(C_l = 0\) and the static Love number vanishes. A reflecting
surface generally sets different inner boundary data and produces
\(C_l\neq 0\).

\bibliographystyle{toine}
\bibliography{biblio}

\end{document}